\documentclass[preprint,12pt]{elsarticle}

\usepackage{amssymb}
\usepackage{graphicx}
\usepackage{color}
\usepackage{subfigure}

\newenvironment{definition}[1][Definition]{\begin{trivlist}
\item[\hskip \labelsep {\bfseries #1}]}{\end{trivlist}}

\begin{document}
\begin{frontmatter}



\title{Ranking the Importance of Nodes of Complex Networks by the Equivalence Classes Approach}


\author[1,2,3]{Bojin Zheng}
\author[3]{Deyi Li}
\author[3]{Guisheng Chen}
\author[1]{Wenhua Du}
\author[1]{Jianmin Wang}
\address[1]{College of Computer Science, South-Central University for Nationalities, Wuhan 430074, China}
\address[2]{State Key Laboratory of  Networking and Switching Technology, Beijing University of Posts and  Telecommunications, Beijing 100876, China}
\address[3]{School of Software, Tsinghua University, Beijing 100084,China}

\begin{abstract}
Identifying the importance of nodes of complex networks is of interest to the research of Social Networks, Biological Networks etc.. Current researchers have proposed several measures or algorithms, such as betweenness, PageRank and HITS etc., to identify the node importance. However, these measures are based on different aspects of properties of nodes, and often conflict with the others. A reasonable, fair standard is needed for evaluating and comparing these algorithms. This paper develops a framework as the standard for ranking the importance of nodes. Four intuitive rules are suggested to measure the node importance, and the equivalence classes approach is employed to resolve the conflicts and aggregate the results of the rules. To quantitatively compare the algorithms, the performance indicators are also proposed based on a similarity measure. Three widely used real-world networks are used as the test-beds. The experimental results illustrate the feasibility of this framework and show that both algorithms, PageRank and HITS, perform well with bias when dealing with the tested networks. Furthermore, this paper uses the proposed approach to analyze the structure of the Internet, and draws out the kernel of the Internet with dense links.
\end{abstract}


\end{frontmatter}

\section{Introduction}

Since Strogatz  and Watts \cite{16} found the small-world effect and Barab\'{a}si and Albert
\cite{1} found the scale-free property of the World Wide Web,  researches on complex networks have greatly increased. These pioneering efforts sought to find some universal principles and invariants, but when people encounter a specific complex network, some questions often  arise: ``Can we identify the most important
nodes in the network?'';``Are these nodes more important than other nodes?'' etc.. These problems are related to data analysis of complex networks. In this field, the node importance is a foundational problem. Only when we know how to measure the importance of nodes, can we answer the questions mentioned above.

The problem of the node importance appears in various fields. For example, studies of the most important scientists\cite{58,70},
the most dangerous terrorists\cite{75,76}, the most critical proteins or genes\cite{18,a23}, etc.. In some circumstances,
we may also need to know the relative important scientists, the relative dangerous terrorists, the relative
critical proteins, etc..
Besides, even if some real-world complex networks are not very large and with the help of computer visualization technologies, people can not yet understand their holistic features. For example, the Internet\cite{221,222}.
We need to find some subnetworks which are combined with some special nodes to obtain holistic understanding of the whole network\cite{152,171}.

``The importance of nodes'' is a vague concept, although many academic papers use it to describe the properties of networks. Researchers have reached little consensus on this concept, but fortunately most concede that it can indeed be described by some rules or the intuitive ideas. As a reference, studies of social
networks, which are certainly regarded as a kind of complex networks, inspired us. In the field of
social networks, the importance of nodes is related to ``centrality''\cite{17}. Moreover, in the field of Web
search engines, the importance of nodes often relies on the importance of neighborhoods, such as in
PageRank\cite{22,23} and HITS\cite{24}. Some papers also propose new ideas on the definition of importance,
such as the failure of system owing to the deletion of one node\cite{60,59}. Considering that all these definitions are very
different, when people deal with some specific applications, they may choose a subset
of these definitions. Based on the investigations on these definitions, we suggest that a set of four rules can be
used to characterize the node importance in common circumstances. Of course, the rules can be inserted or removed as needed to accommodate the particular real world situations.

However, these rules often conflict to the others. We need a solution to resolve the conflicts.

Mathematically, each of the four rules defines an order relationship on the importance of nodes. For example, if we define the important nodes as those with more neighbors (larger degree value) then the degree value actually is a measure of the node importance. Because the node importance needs more than one rule to characterize, when we have
characterized the concept and chosen corresponding rules and formulas, we need to focus on how to deal with
the conflicts among these order relationships and aggregate the results of the rules. Usually, the aggregation
should not have bias on any rule. 

There are two relative works on the aggregation. In the field of Web search engines, people use a smart method named ``Rank
Aggregation''\cite{64} to deal with this problem. In another way, if we treat every rule as an
optimization object, this problem can be transferred to a multi-objective optimization problem. In the field
of Evolutionary Algorithms for multi-objective optimization problems, this is commonly tackled with a
mathematical tool, called the dominance relationship.
The dominance relationship can categorize the nodes into the equivalence classes.
Every
equivalence class will indicate the same ordinal number of the nodes in it. The first equivalence class would
only include the most important nodes, which are also called ``the skyline''\cite{a3,a4} in the field of Database Management System (DBMS) and ``the Pareto front''\cite{66} in the field of Evolutionary Computation. 

Here we suggest that the equivalence classes can be used to resolve the conflicts among the indicators based on the following reasons: 1) this approach has no bias on any rule; 2) this approach can obtain diverse and representative important nodes; 3) this approach can guarantee that a good node would certainly have a good ordinal, which is very important to the explanation of the ordered results; 4) this approach has a mathematical foundation, i.e., it is related to the maximum vector problem\cite{a1}; 5) this approach has been used in various fields, such as DBMS and Evolutionary Computation.

Because PageRank and HITS are very successful in Web Searching, some researchers believe that they will
perform well in ranking the nodes of complex networks and then use them as benchmark algorithms to measure the effectiveness of the other algorithms
\cite{25,55,77,78}, but whether they can perform well or not when applying them for very
different purposes has not been discussed before. Though PageRank and HITS are perfect under their assumptions,
the impossibility theorems restrict their extensions\cite{73,74}, so that they cannot be taken for granted as benchmarks. Assume that we set the intuitive rules as the benchmark, thus, are the results of PageRank and HITS similar to the rules of the node importance here? Considering that both the results of the rules and the results of these algorithms can be represented as sequences, we suggest the Kendall's $\tau$\cite{72} as the measure indicator. Based on this idea, we define a measure indicator and three sub-indicators, and introduce an
algorithm to calculate them. These sub-indicators can be regarded as the effectiveness of the ranking algorithms relative to the framework.

To validate our ideas, we chose three well-known real-world networks, i.e., the metabolic network\cite{18}, the
dolphins network\cite{20} and the Zachary karate club network\cite{19} as the examples to carry out the
experiments. The experimental results show that the framework can produce representative and diverse nodes. Moreover, we also calculate the
effectiveness measure sub-indicators. The experimental results show that PageRank and HITS perform well even though they were not designed for the undirected networks. However, the results also show that both algorithms have bias.

In general, our work tries to provide a framework for issues on ranking/sorting the node importance. This framework addresses three main issues. First, how to reasonably define the node importance? Our solution is to clarify the concept of the node importance from the intuition by analyzing
the concept with rules that define every aspect of this concept respectively.  Second, how to resolve
conflicts among rules? Our solution is to aggregate the results of the rules by importing a mathematical tool
that categorizes them into the equivalence classes and then assemble them into a partially ordered sequence. Third, how
to measure the effectiveness of the compared algorithms? Our solution is to compute the similarity between the results of the framework and that of the compared algorithm. We uses three widely used networks to validate the approach. Finally, we apply the proposed approach to analyze the structure of the Internet.

\section{Methods}

\subsection{The problems in the node importance}

The importance of nodes is related to many fields, and it is currently attracting more and more interests.

The evaluation of the node importance is based on the researches in Graph Theory and Graph-based Data
Mining\cite{61,86,91}, and we can say that this field can be regarded as a branch of Graph-based Data Mining. However, research on the node importance origins from the study of social networks. After the emergence of complex networks, other applications in technical networks, biological networks, etc. were also proposed.

To clarify the node importance, we need to solve three main problems. The first one is that what are the suitable rules to define the node importance. The second one is how to resolve the conflicts among all the rules. The third one is how to measure the effectiveness.

\subsubsection{Defining the importance of nodes}

The importance of nodes was discussed in social networks as ``centrality''. Before the work of Freeman, there
is no unanimity on what is ``centrality'' and what is the conceptual foundation of ``centrality''; Of course, there is no unanimity on what is the proper measure of ``centrality''\cite{17}.

In 1976, Freeman reviewed the measure of ``centrality''\cite{17} and suggested three measure indicators for
point (node) centrality. First, ``degree'' is a proper measure, it can be used to measure the communication activity
in a network. Second, ``betweenness'' is based upon the frequency with which a node falls between pairs of other
nodes on the shortest or geodesic paths connecting them, and is used to exhibit a potential for control of
the communication. Third, ``closeness'' is based upon the degree to which a node is close (approximates) to all other nodes in the
network and is used to exhibit the independence or efficiency of communication. These three measure indicators
are referenced as three different structural attributes.

The work of Freeman has achieved a great success. Many researchers use the ``centrality'' under Freeman's
suggestion now.

The importance of nodes was defined recursively in PageRank and HITS when dealing with the Web documents,
that is, the importance of a node relies on the importance of neighbors, and the neighbors' relies on
neighbors' neighbors', etc.. Essentially, the importance of nodes is related to the degree of nodes. Based on this
idea, PageRank and HITS need to iteratively calculate the node importance. However, both these algorithms
were designed to deal with directed graphs. Because this paper mainly focuses on undirected graphs, the
undirected vertex are treated as a couple of vertexes with opposite directions to make both the algorithms feasible.

The importance of nodes also was defined as the failure of system\cite{59}. If a node is removed from a
connected graph, and the new graph is not connected, then this node is important. Actually, literature\cite{59}
and the related papers emphasize on the importance of node sets. For a single node, this definition may
include less information because the nodes mostly do not break the network only by itself. Moreover, this definition can be partially represented by the betweenness.

Moreover, when considering the spreading process on complex networks, the K-shell\cite{a22} can be used to define the most central, or the most important nodes.

Notice that this paper focuses on the static structure, the K-shell and the importance definition on failure\cite{59} are not applicable. Moreover, the definitions of PageRank and HITS can be represented as the neighborhood importance, therefore, four rules among these definitions are suggested to measure the node importance.

\subsubsection{Resolving the Conflicts}
If we regard every rule as an optimization object, the problem how to resolve the conflict among the rules could be
transferred to a multi-objective optimization problem. In the field of multi-objective optimization,
the simplest way is to set weights\cite{65}, which reflects the users' preference for every rule. Because the
weights in the Weighted Sum Method are uncertain, this method cannot be used for a benchmark purpose.

Maybe we can refer to a smart technology named ``Rank Aggregation''\cite{64} in Web search engine. The aim of the
rank aggregation is to obtain a sequence which has a minimal distance to all ranking results from different
measure indicators or ranking algorithms. If we use this method, that means we actually assume that the sequence with minimal distance would be most satisfactory to the definition of the node importance. This assumption is not necessary. Moreover, this method would not guarantee that the good node has a good ordinal.
For example, if node $a$ is more important than node $b$ under all the rules, the rank aggregation may set that node $b$ is more important than node $a$ to minimize the distance.

In the field of multi-objective optimization, ``dominance relationship'' is used to deal with the
conflicts\cite{81,66,67,68}. In contrast to the Weighted Sum Method, this method does not need parameters,
and moreover, it is not sensitive to the small difference of scores of nodes, because it is based on order
relationships. This method can keep the diversity to obtain representative important nodes. For example,
the best node under every rule should certainly be regarded as one of the most important nodes. Furthermore,
when using it as a benchmark, if node $a$ is better than node $b$ in all aspects, node $a$ would certainly
be more important than node $b$. This feature is useful for a reasonable explanation.

Moreover, the concept of ``Pareto front'' is used to describe the set of ``the
most important nodes''. Cotta and Merolo\cite{58,70} referred to the work of scientific collaboration network
\cite{80} and computed the degree, the betweenness and the closeness of the scholars majoring in Evolutionary
Computation and obtained interesting results. They also listed the ``Pareto
front''--the most important scholars. In this paper, we extend this method and used it as a good benchmark, and
formally describe the detailed mechanism and introduce how to pack the results into a sequence and how to
compare to other algorithms.

\subsubsection{Measuring the effectiveness}
When we have obtained the results from the benchmark and compared algorithm, how to define the effectiveness
measure indicator(s) is an important problem. Because both the results can be expressed as the sequences, the
effectiveness actually is the ``distance'' between two sequences.  Two popular measures are related to this
issue: the Spearman's $\rho$ and the Kendall's $\tau$\cite{72}. Kendall's $\tau$ is based on such an idea: how many
times to transfer a sequence to the other by neighbor swapping.

When applying Kendall's $\tau$ to compute the effectiveness, the compared algorithms can ``cheat'' the measure
indicator by generating too many equivalent nodes, so we use two sub-indicators to bound it and a sub-indicator to
indicate the degree of ``cheating''.

Here, we have introduced the problems, then we will introduce more detailed information on the solutions in the next subsections.

\subsection{Obtaining the rules from intuition}
People reached little consensus on the concept of ``the importance of nodes'', but most concede
that the node importance can be described by intuitive ideas. For example, ``degree'' is used as a measure indicator for
the communication activity in human communication. A person with more social relationships is treated as more
important; and in a pure network, from one aspect, the nodes with larger degree are regarded as more
important. This description on the node importance in networks is obviously based on intuition. Therefore,
when we discuss the node importance, we must extract the formal representation from the intuition.

\subsubsection{Rules and the selection of rules}

We choose four rules to evaluate the node importance according to the existing researches.
\begin{enumerate}
  \item [Rule] 1: Commonly, the nodes with larger degree would have a larger influence on the network, so we
  think that the nodes with larger degree are more important.
That is, if the degree of node $a$ is larger than that of node $b$, then node $a$ is more important than node $b$.

  \item [Rule] 2:  If a node connects two or more communities, it is a key node, because this node can control the communication
between the communities. We know, the potential of controlling the communication can be measured by betweenness.
As such, if the betweenness of node $a$ is larger than that of node $b$, then node $a$ is more important than node $b$.

  \item [Rule] 3:  If a node is closer to ``the center'' of the network, it is more important.
  Because closeness is the proper
indicator, we have, if the closeness of node $a$ is larger than that of node $b$,
then node $a$ is more important than node $b$.

  \item [Rule] 4:  If a node has neighbors with larger influence, then this node is more important.
  That is, if the neighbors
of node $a$ have larger influence than that of node $b$, then node $a$ is more important than node $b$.
\end{enumerate}

The intuition of rules 1, 2 and 3 come from the field of social networks, and the intuition of rule 4 from the Web
search engine. 

\subsubsection{Formalizing the four rules}
Though we list four rules to describe the concept of the node importance, mathematical
formalization still is necessary.

We denote ``the importance of node $a$ is greater than that of node $b$'' as $a > b$, and ``the
importance of node $a$ equals that of node $b$'' as $a = b$.

According to the definitions above, the rules can be formalized as follows,

Rule 1 involves the degree as the measure indicator. We note
the degree of node $a$ as $Degree(a)$.
\begin{equation}\label{equ.rule.2}
Degree(a)= \sum {\delta (i,a)}
\end{equation}
where,

\begin{equation}\label{equ.rule.2.1}
\delta (i,a) = \left\{ \begin{array}{l}
 1,{\rm{if\ node\ i\ is\ connected\ to\ node\ a}} \\
 0,{\rm{if\ node\ i\ is\ NOT\ connected\ to\ node\ a}} \\
 \end{array} \right.
\end{equation}

Rule 2 can be measured by betweenness, $Betweenness(a)$ would represent the betweenness of node
$a$ and is defined as follows,

Assume that $B(i,j)$ is the number of shortest paths (geodesic paths) between node $i$ and $j$, and have
$B(i,j)>0$, $B(i,a,j)$ is the number of shortest paths between node $i$ and $j$ and through the
node $a$. Then we have,

\begin{equation}\label{equ.rule.3}
Betweenness(a) = \sum\limits_{i \ne j} {\frac{{B(i,a,j)}}{{B(i,j)}}}
\end{equation}

Rule 3 actually is based upon the closeness, we use $Closeness(a)$  to represent the closeness measure of node
$a$, $d_{ij}$ to represent the length of the shortest paths between node $i$ and $j$ and $V$ to
represent the set of all nodes in the network. Then we have,

\begin{equation}\label{equ.rule.4}
Closeness(i ) = \frac{1}{{\sum\limits_{v_j  \in V \wedge i \ne j} {d_{ij} } }}
\end{equation}

As to rule 4, we use a function
of degrees of the neighbors of the node to measure it. We use $Nb(a)$ to represent the set of the immediate
neighbors of node $a$, that is, all elements in $Nb(a)$ would have and only have a hop/transimition from node $a$.

And we suggest such a function,
\begin{equation}\label{equ.rule.5}
Neighbors(a) = \sqrt[nk]{{\sum\limits_{i \in Nb(a)} {Degree(i)^{nk} } }}
\end{equation}
If $nk>1$ then the neighboring nodes with larger degree would contribute more to the value of this
indicator, but this may make a dilemma that the nodes with larger degree may be inferior to its
neighbors with larger degree under this rule. For example, when the network structure is a star or a wheel, the
central nodes would have an inferior indicator value. And v.s., if $nk<1$, then the neighboring nodes with larger degree
would contribute less, which is unfair, unless for some specific purposes. So we suggest $nk=1$.

Based on the definitions above, the rules would be depicted as follows,
\begin{enumerate}
  \item [Rule] 1:$N\setminus{\{a\}}\equiv N\setminus{\{b\}} \Rightarrow a=b $
  \item [Rule] 2:$Degree(a) >= Degree(b)\Rightarrow a >= b$
  \item [Rule] 3:$Betweenness(a) >= Betweenness(b)\Rightarrow a >= b$
  \item [Rule] 4:$Closeness(a) >= Closeness(b) \Rightarrow a >= b$
  \item [Rule] 5:$Neighbors(a) >= Neighbors(b) \Rightarrow a >= b$
\end{enumerate}

\subsection{Ordering the nodes by the dominance relationship}
Four rules tell us the order relationships under different measure aspects. Because they may be
conflicting, we need a method to aggregate their results. Here we employ a kind of
strict partial order relationship called dominance relationship to order the nodes.

\subsubsection{Basic ideas}
\begin{figure}[!ht]
\begin{center}
  \includegraphics[width=7.2cm,height=5.4cm]{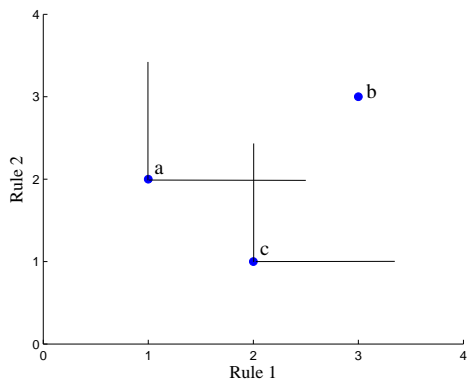}\\
  \end{center}
  \caption{{\bf The Illustration of Dominance Relationship.} The more important the node is, the smaller its ranking number is. From this figure, node a and node b dominate node c, because that both nodes are more important than node c, no matter with the rule 1 or the rule 2. Node a can not dominate node b, and vice versa. }\label{fig.ranking.sketch}
\end{figure}

For an illustration, we assume only two rules 1,2 and three nodes $a$, $b$, $c$. Assume that under rule 1, $a > c
> b$; and under rule 2, $c > a > b$; Then, for rule 1, we set values of $a$, $c$, $b$ as 1, 2, 3; for rule 2, we
set values of $c$, $a$, $b$  as 1, 2, 3. So $a$, $b$, $c$ would have coordinates $(1,2) ,(3,3),(2,1) $. Here, smaller
number represents more importance. See Fig. \ref{fig.ranking.sketch}.

Because node $a$ is more important than $b$ for rule 1 and rule 2, so we say node $a$
\textbf{\emph{dominates}} node $b$ or, $b$ \textbf{\emph{is dominated}} by node $a$. Obviously, $a$
and $c$ are not dominated by any other nodes, so we say they rank 1. If we erase all non-dominated
nodes, that is, node $a$ and node $c$, from Fig. \ref{fig.ranking.sketch}, then node $b$ would not
be dominated by any other nodes, so we say node $b$ ranks 2.

For each node, it is assigned four values to represent the ordinals under the corresponding rules. That is, the node importance can be represented by a vector with four elements. For node $a$, we denote
$a(a_1,a_2,a_3,a_4)$, here $a_1$ represents the ordinal value under rule 1, $a_2$ represents the ordinal value
under rule 2 and so on.

Therefore, rule 1 to rule 4 can be rewritten as follows,
\begin{enumerate}
  \item [Rule] 1:$Degree(a) >= Degree(b)\Rightarrow a_1 <= b_1$
  \item [Rule] 2:$Betweenness(a) >= Betweenness(b)\Rightarrow a_2 <= b_2$
  \item [Rule] 3:$Closeness(a) >= Closeness(b) \Rightarrow a_3 <= b_3 $
  \item [Rule] 4:$Neighbors(a) >= Neighbors(b) \Rightarrow a_4 <= b_4$
\end{enumerate}

According to the rules, the set of the importance vector of all nodes is a strict partial order set
based on the dominance relationship and satisfies the irreflexive, transitive, asymmetric relationships. Because the dominance relationships are a kind of strict partial order
relationship, and a strict partial order relationship should respond to a non-strict partial order relationship\footnote{A non-strict partial order relationship is a binary relationship ``$\leq$'' over a set which is reflexive, antisymmetric, and transitive; A strict partial order relationship ``$<$'' is a binary relation over a set which is irreflexive and transitive, and therefore asymmetric.}.
Under the non-strict partial order relationship, the nodes can be packed into the equivalence classes, that is, all the nodes in the same equivalence class are equivalent in certain relationship.

We give formal
definition for the dominance relationship below,

\begin{definition}[Dominance] Given node $a$ and node $b$, $a(a_1,a_2,\ldots,a_m)$ (m is the number of the selected
rules) is said to dominate
$b(b_1,b_2,\ldots,b_m)$ (denoted by $a \prec b$), if and only if
$a$ is less than or equivalent to $b$ for each element and $a \neq b$, i.e.\\
\begin{equation}\label{equ.dominance}
    \forall i \in \{1,2,\ldots ,m\}, a_{i} \le b_{i} \wedge \exists i \in \{1,2,\ldots ,m\}, a_{i} <
b_{i}
\end{equation}
\end{definition}

\begin{definition}[Non-dominated Set] Given a set A, the Non-dominated Set (denoted as NDS) is the set that contains all the nodes which are
not dominated by any other nodes. Mathematically, \textit{NDS} is defined as:
\begin{equation}\label{eqn.nds}
 \textit{NDS}= \{u \vert \nexists x \in A: x \prec u \}
\end{equation}
\end{definition}

Here NDS is ``the Pareto front'', or ``the skyline set''\cite{a3}. Mathematically, NDS is the maxima of a set of vectors\cite{a1}.

Given a set A, the elements in $NDS$ would rank 1. $(A - NDS)$ is the dominated set, and if it is not empty, it would have its NDS, all the elements in $(A - NDS)$'s NDS would rank 2. Every NDS in different hierarchies is an equivalence class.

When we set the ordinal to every node, we can categorize all the nodes into the equivalence classes based on
iterative non-dominated sets, and every equivalence class means a ranking number.

Therefore, the procedure to rank the nodes by the importance actually is the procedure to
iteratively obtain the non-dominated set from the remaining set which has eliminated the
non-dominated nodes.

\subsubsection{The algorithm of the equivalence classes}
Once we determine the measure indicators of the importance, then we have determined the results of the equivalence classes based on the dominance relationship, that is, the iterative non-dominated sets.
So we need an algorithm to compute the equivalence classes.

In the field of DBMS, several algorithms have been proposed to compute the skyline set efficiently. For examples, Block-Nested-Loops algorithm, Divide-and-Conquer algorithm and B-Tree algorithm proposed by B\"{o}rzs\"{o}nyi et al.\cite{a3}, the progressive skyline algorithms proposed by Tan et al.\cite{a4} and the online algorithm proposed by Kossmann et al.\cite{a5}.

These algorithms can also be used to compute the equivalence classes efficiently by iteratively computing the skyline set. Here we just introduce a simple algorithm, regardless of efficiency, to categorize the nodes into the equivalence classes. Assume that the input is a
set named $S$, we define a set named $NDS$ to store the non-dominated nodes, and a set named
$Extra$ to store dominated nodes in current iteration step. When an iteration step is executed, the equivalence
classes and the rank numbers of the equivalence classes can be output step by step.

\begin{figure}[!ht]
\begin{center}
  \includegraphics[width=9.39cm,height=6.03cm]{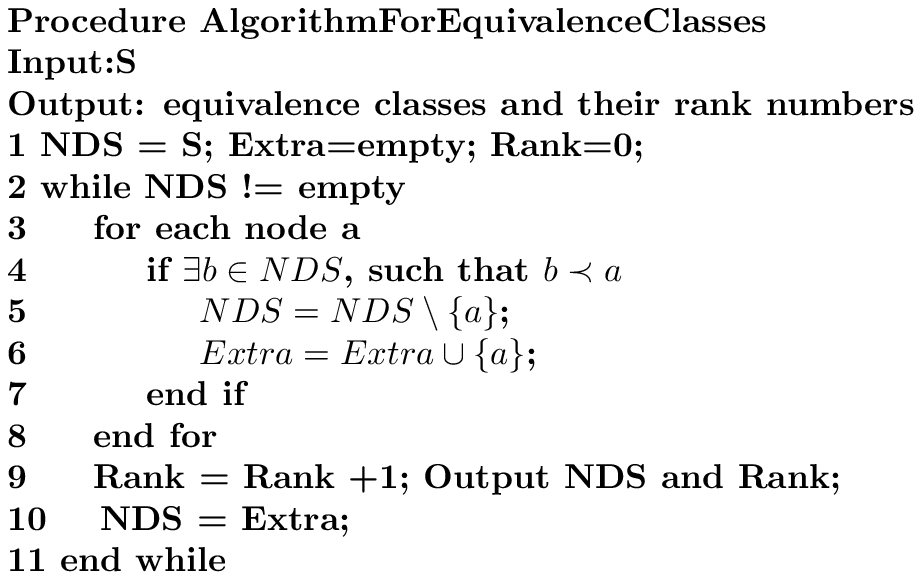}\\
  \end{center}
  \caption{{\bf The Pseudocode for the Equivalence Classes.} This code illustrates the algorithm to obtain the equivalence classes. It actually is an iterating process to extract the outermost nodes from the whole set.  }\label{fig.pseudocode.ec}
\end{figure}

The pseudocode of this algorithm is depicted as Fig. \ref{fig.pseudocode.ec}.

\subsection{Measuring the effectiveness of ranking/Sorting algorithms}
Once the rules are determined, the nodes could be categorized into the equivalence classes by their
importance. The nodes in the same equivalence class will share the same rank number, so all the nodes could be packed into a partially ordered sequence.

But the further problem would be, because several algorithms have been used to sort/rank the nodes by the importance,
how to measure the effectiveness of algorithms? i.e., how to measure the similarity between the
aggregation of the results of the selected rules from intuition and the results of the ranking algorithms?

PageRank\cite{22,23} and HITS\cite{24} are the most famous algorithms and they were designed to
automatically determine the importance of Web pages. Here we use them 
for the analysis of complex networks like the previous works\cite{25}, and we try to investigate how much PageRank and HITS are similar to the equivalence classes approach.

PageRank and HITS are the ``global ranking/sorting'', that is, it will generate a totally ordered sequence. A totally ordered sequence means that every element in this sequence can be ordered by a property and has an identical ordinal. Because both the results of the rules and the results of the compared algorithm can be expressed as sequences, we can use the distance between the partially ordered sequence generated by the
equivalence classes approach and the totally ordered sequence generated by the compared algorithms to measure
the effectiveness of the algorithms.

If we use the partially ordered sequence as a benchmark for comparison, every opposition to the
partially ordered sequence is regarded as a unit of distance. According to this idea, the distance can be defined.  Actually, the distance is an indicator of the similarity. When we know the maximum of distance, we can define the similarity ratio.

We use the term ``coverage'' to represent the similarity ratio.

\subsubsection{coverage}

We use a short example to illustrate the concept of coverage. Assume that there is a partially
ordered sequence, denoted as $\{[1,2],[3,4]\}$ which means $\{1,2\}$ is an equivalence class,
and $\{3,4\}$ is another, and a totally ordered sequence, $\{4, 3, 2, 1\}$ which means $4 > 3 > 2
> 1$. Then how to measure the distance between two sequences? We use the counting of swapping
between the neighboring nodes to define the distance, that is, how many times we swap the neighbors to transfer
the totally ordered sequence to meet the partially ordered sequence, the least number of actions can be defined as the distance. As to
the example, we need at least 4 times, that is, $3 \leftrightarrow 2,3 \leftrightarrow 1, 4 \leftrightarrow
2, 4 \leftrightarrow 1$; therefore, the distance is 4. Actually, the distance means the violation of the
order relationships to the partially ordered sequence. Notice that the position relationship between $1$ and $2$ (similar for $3$ and $4$) is not considered.

Once the distance is defined, the coverage can be
defined as the opposition of the ratio of violation. As to the example, because 4 is the possible
largest value for all possible totally ordered sequence, so the coverage is zero.

In many
circumstances, the algorithms would estimate some inequivalent nodes as equivalent, so we define two sub-indicators --
the best and worst coverage -- to evaluate the coverage in the best and worst circumstances, that is, if we
want to get the worst coverage and the compared algorithms obtain a sequence $\{4, [2, 3], 1\}$, that means no. 3
node equals to no. 2 node, because for the partially ordered sequence, no. 2 node is prior to no. 3 node, so we
rewrite it as $\{4, [3, 2], 1\}$, such that the relative position of no. 2 node and no. 3 node is different
from that of the partially ordered sequence, here, the coverage is the worst coverage.

Here we formally depicted the concept of coverage as follows,

\begin{definition}[neighbor swapping] Given two sequences $a(a_1,a_2,...,a_n)$ and $b(b_1,b_2,...,b_n)$,
if $\exists i \in [1,n]$ such that $a_i = b_{i+1}$ and $a_{i+1} = b_{i}$ and for arbitrary $j, j
\neq i \wedge j \neq i + 1, a_j= b_j $, then we say, there exists one time of neighbor swapping to
transfer the sequence $a$ to $b$, denoted $NS(a,b) =1$.
\end{definition}

\begin{definition}[distance between two sequences] Given two sequences $a$
and $b$, if there exist the least k sequences $s_1, s_2,..., s_k$ such that $\forall j\in [1,k-1],
NS(s_j,s_{j+1})=1 \wedge  NS(a,s_1)=1 \wedge NS(s_k,b)=1$, then we say $NS(a,b) = k+1$ and the
distance from $a$ to $b$ is $dis(a,b)=k+1$.
\end{definition}

A partially ordered sequence can actually be regarded as a set of sequences, so
the distance between a sequence
and a partially ordered sequence is the minimum distance from a sequence to a set of
sequences.

\begin{definition}[distance from a sequence to a sequence set] Given a sequence $a$
and a sequence set $S$, the distance from $a$ to $S$ is $dis(a,S)= min\{dis(a,i) \mid i \in S\}$.
\end{definition}

For a given network, assume that the partially ordered sequence of the equivalence classes approach is $P$, and the compared sequence is $s$,
and all the possible sequences of $P$ would construct a set denoted $PS$. Then the coverage can be depicted as
follows,
\begin{equation}\label{eqn.coverage}
    coverage(s) = 1 - \frac{{dis(s,P)}}{{Max(\{ dis(k,P){\rm{|}}k \in PS\} )}}
\end{equation}

When the sequences of the compared algorithms are totally ordered sequences, this indicator is enough to characterize the similarity between these sequences to the sequence of the equivalence classes approach. But some algorithms would generate some equivalent nodes, then, how to define the similarity ratio?

As to the sequences with some equivalent nodes,
we can regard it as a sequence set and choose the
best sequence and the worst sequence in the sequence set to represent it, therefore, we define the best
coverage and the worst coverage.

If we regard the compared sequence with the equivalent nodes as a set of sequences (denoted $S$), we have the
definition of two sub-indicators, best coverage and worst coverage, based on coverage as follows,

\begin{equation}\label{eqn.bestcoverage}
    Bestcoverage(s) = 1 - \frac{{Min(dis(s,P)|s \in S)}}{{Max(\{ dis(k,P){\rm{|}}k \in PS\} )}}
\end{equation}

Notice that the best coverage does not measure the best effectiveness, because when all the nodes are in an
equivalence class, the best coverage is 1, here it is meaningless, because the worst coverage is 0.
The worst coverage is a more proper measure indicator to measure the worst effectiveness. And the gap
between the best coverage and the worst coverage would mean the number of nodes which are incorrectly
regarded as equivalent nodes. Here
we define other sub-indicator ``certainty ratio'' that are denoted as ``certratio'' to indicate the incorrect degree.

If the compared algorithm can identify the equal nodes and does
not regard the inequivalent nodes as equivalent, its certratio is 0.  Actually, if the algorithm treats every
nodes uniformly, and does not import extra information, the symmetric nodes will not be assigned different
ranking values. Under such an assumption, the certratio is only related to the gap between the best coverage and
the worst coverage. Here we define the certratio as equation \ref{eqn.certratio}, which is not suitable for
some special algorithms which include extra information.

\begin{equation}\label{eqn.certratio}
    certratio(s) = bestcoverage(s) - worstcoverage(s)
\end{equation}

\subsubsection{Algorithm to calculate coverage}
The calculation of coverage includes two steps. The first step is the calculation of
the distance between $s$ and $P$. The minimum neighbor swapping seemly is difficult to be counted, but there
is an easy algorithm to solve this problem.

Every node would have two property values. One is the rank number, the other is the importance value
assigned by the compared algorithms. If we sort the nodes by the importance value, then the
distance is the counting of neighbor swapping to sort the nodes by the rank number.

For example, if there are four nodes $a, b, c, d$, the rank numbers are $1, 1, 2, 3$. Assume that the
compared algorithm has generated a sequence, that is, $d, c, b, a$, so the corresponding rank number
sequence is $3, 2, 1, 1$, the minimum counting of neighbor swapping is the counting of
neighbor swapping to ascendingly sort the rank number sequence.

\begin{figure}[!ht]
\begin{center}
  \includegraphics[width=11.75cm,height=5.19cm]{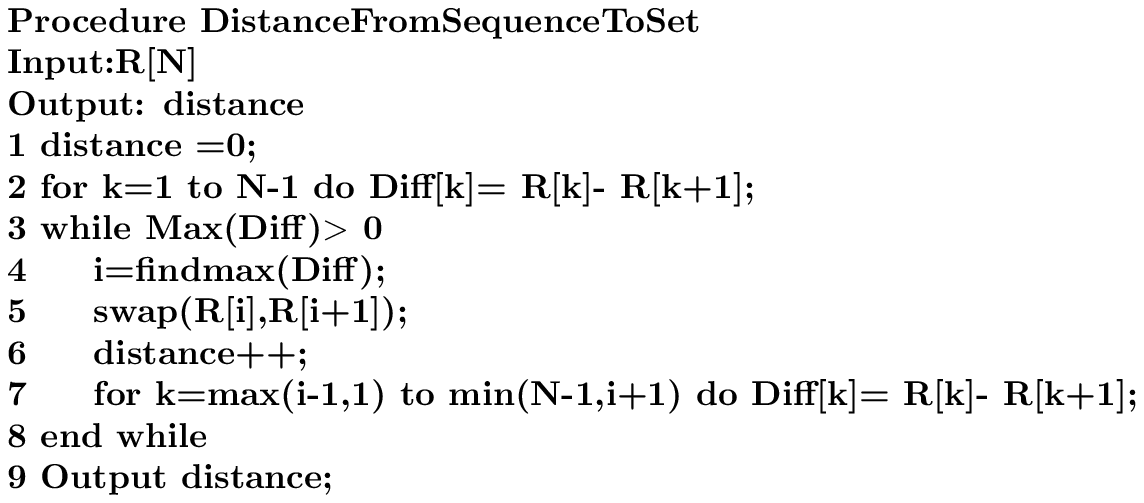}\\
  \end{center}
  \caption{{\bf The Pseudocode for Calculating the Distance.} This Pseudocode illustrates the algorithm to compute the distance from a sequence to a set. The distance is defined as the smallest number to move a sequence to match the set. }\label{fig.pseudocode.distance}
\end{figure}

To obtain the minimum of neighbor swapping, no neighbor swapping could be duplicated. We use the
greedy strategy to deal with the problem. That is, we define an array which records the difference
between every node and its successor, then the maximum difference is preferred to be
eliminated, that is, this corresponding node and its successor is firstly swapped. So we use
the array $Diff[N-1]$ (here N is the number of the nodes) to record the difference of the array of the nodes
$R[N]$, and iteratively swap the nodes with maximum difference to its successor. Notice that when
the rank value sequence are sorted, the elements in array $Diff[N-1]$ are -1 or 0, so
the stopping criteria of this algorithm is that the maximum of elements in $Diff[N-1]$ is
smaller than zero. The algorithm can be depicted as Fig. \ref{fig.pseudocode.distance}.

The second step is the calculation of maximum possible distance to the partially ordered sequence. The
intuitive idea is to construct the furthest sequence and count the neighbor swapping. But there is a
simper formula.

Assume that the nodes are categorized into $NE$ equivalence classes denoted as $E_1, E_2,..., E_{NE}$,
and the sizes are $|E_1|,|E_2|,...,|E_{NE}|$. Notice that the maximum distance between arbitrary
nodes is $N(N - 1)/2$, each equivalence class $E$ would result in a subtract of $ |E|(|E|-1)
/2$, so we have,
\begin{equation}\label{eqn.maxdistance}
    maxdistance  = \frac{{N(N - 1)}}{2} - \sum\limits_{i = 1}^{NE} {\frac{{|E_i |(|E_i | - 1)}}{2}}
\end{equation}

Based on the methods above, the best coverage and the worst coverage is easy to calculate, because we only need to deal with the nodes with equivalent importance. Before the calculation of the minimum of the distance, if we sort the nodes with the same importance according to the same orders in the sequences of the equivalence classes approach, then we obtain the best sequence. If according to reverse orders, the we obtain the worst sequence.

\section{Results}
We choose three real-world networks to validate the proposed approach. The selected networks are the
metabolic network, the dolphins network and the Zachary karate club network. These three real-world networks
are widely used as examples.

\subsection{Selected Real-world Networks}

\begin{figure}[!ht]
\begin{center}
  \includegraphics[width=8cm,height=6cm]{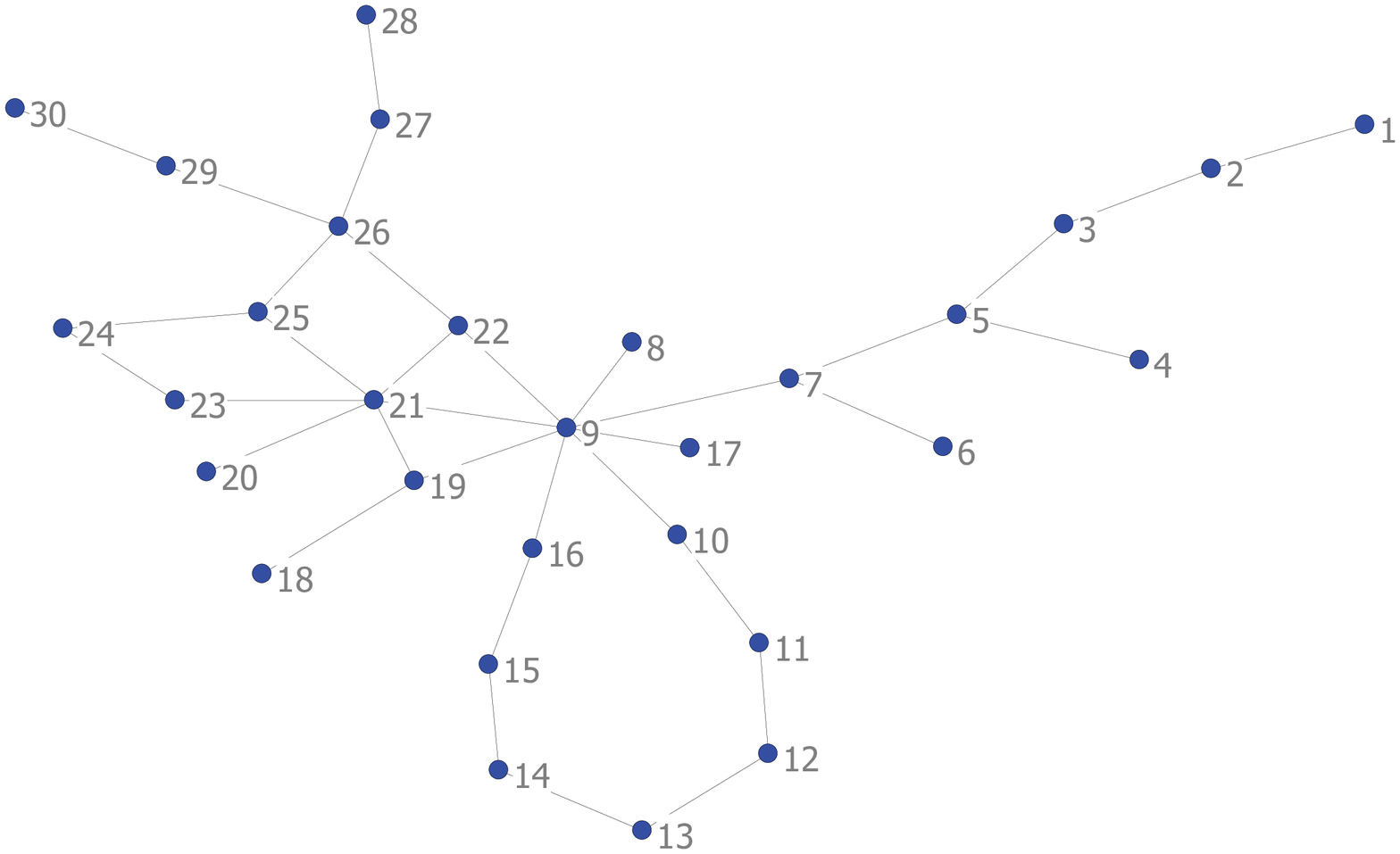}\\
  \end{center}
  \caption{{\bf The metabolic network.} This figure illustrates a module of a metabolic network. Because of its plentiful characteristics and its simpleness, it is widely used as the test-bed example. Among all the nodes, node 9 is the most important.}\label{fig.netdraw.metabolic}
\end{figure}

Protein interaction networks are deeply investigated in biology. The metabolic network is a kind of protein
interaction network. Here we chose a single module in the metabolic network of \emph{A. thaliana},
because the module is simple, but it has many typical features. There are 30 nodes and 34 edges in this network and the edge
represents that two nodes participate in the same metabolic activities. See in Fig. 
\ref{fig.netdraw.metabolic}.

\begin{figure}[!ht]
\begin{center}
  \includegraphics[width=8cm,height=6cm]{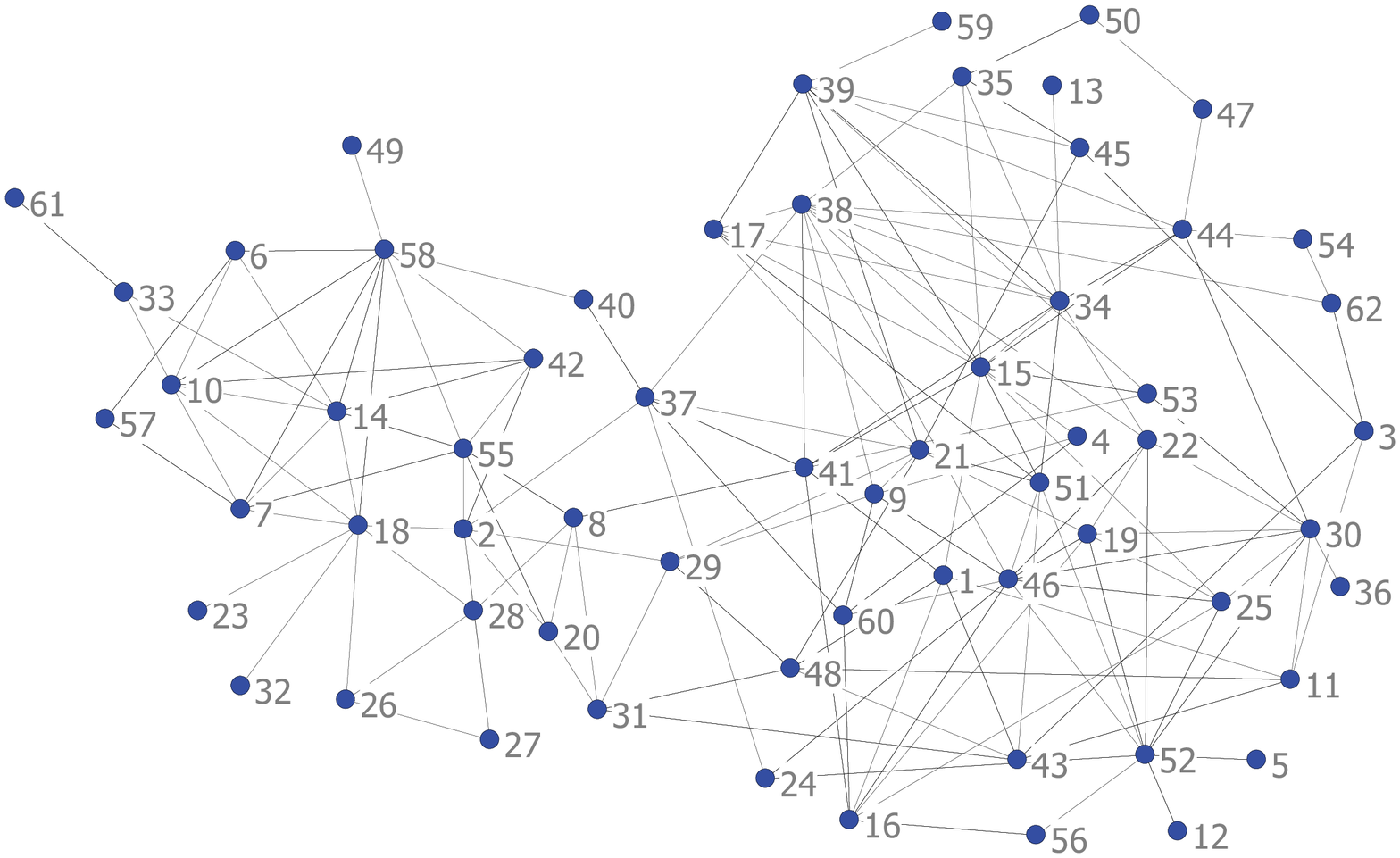}\\
  \end{center}
  \caption{{\bf The dolphins network.} This figure illustrates the relationship network among 76 dolphins. This network has community structure, which are united by node 37. Hence, node 37 is one of the most important node.}\label{fig.netdraw.dolphins}
\end{figure}

The dolphins network owes to the work of D. Lusseau. Dr. D. Lusseau observed a community of 62 bottlenose
dolphins (Tursiops spp.) over a period of 7 years from 1994 to 2001. According to the observation they
constructed the dolphins network.  The nodes represent the dolphins, and the ties between nodes
represent associations between dolphin pairs occurring more often than expected by chance. Node 37 is a
very important node, when node 37 disappeared, the network broke into two communities, one big and one small,
but when it returned, two communities got united. The whole network has 62 nodes and 159 ties. See in Fig.
\ref{fig.netdraw.dolphins}.

\begin{figure}[!ht]
\begin{center}
  \includegraphics[width=8cm,height=6cm]{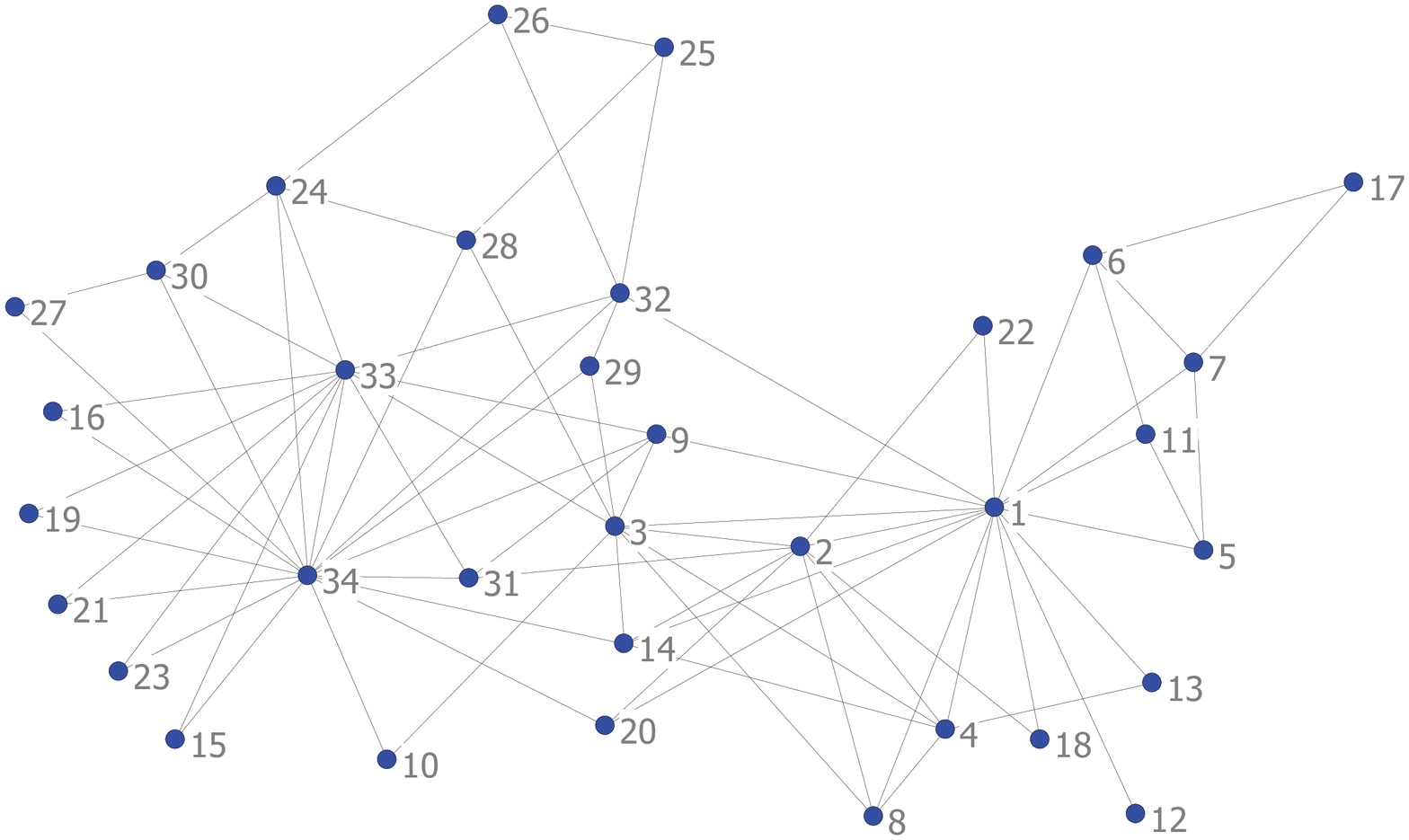}\\
  \end{center}
  \caption{{\bf The Zachary karate club network.} This figure illustrates the relationship network among the members of a karate club. The disagreement among node 1 and node 34 leads to the division, suggesting th most important nodes are node 1 and node 34.}\label{fig.netdraw.zachary}
\end{figure}

The Zachary karate club network is a famous social network. Wayne Zachary observed a karate club in a
university for a period of three years, from 1970 to 1972. In addition to direct observation, the history of
the club prior to the period of the study was reconstructed through informants and club records in the
university archives. The nodes represent the members of the club, and the edges represent the friendship
between the nodes. The 34th node, John A. is the chief administrator and the 1st node, Mr. Hi is the karate
instructor. Because of the conflicts between Mr. Hi and Mr. John, the network broke into two clubs. This
network includes 34 nodes and 78 edges. See in Fig. \ref{fig.netdraw.zachary}.

\subsection{The Equivalence classes of the Tested Networks}
Based on the proposed approach, we calculated the equivalence classes of the tested networks. Here we set $nk = 1$.
Besides, we also computed the results of the measure indicators of degree, betweenness, closeness with the tool
UCINET\cite{79}.

The equivalence classes of the metabolic network are listed in Table \ref{tab.po.metabolic}.

\begin{table}[!ht]
\caption{The equivalence classes of the metabolic network}
\centering \(\begin{array}{|p{30pt}|p{80pt}|| p{30pt}|p{80pt}|} 
\hline Rank&Nodes& Rank & Nodes\\
\hline 1&9 & 2&7 21 26 \\
\hline 3&5 22 & 4&10 16 19 \\
\hline 5&3 11 15 25 27 29 & 6&2 8 12 14 17 23 \\
\hline 7&13 20 24 & 8&6 \\
\hline 9&18 & 10&4 \\
\hline 11&28 30 & 12&1 \\
\hline
\end{array}\)
\label{tab.po.metabolic}
\end{table}

From Table \ref{tab.po.metabolic}, we can see that no. 9 node belongs to the first class of nodes, because no. 9 node has better values of degree, betweenness,
closeness and neighbors indicators. No. 9 node undoubtedly is the most important protein.

According to the same approach, we list the equivalence classes of the dolphins network and the Zachary karate
club network in Table \ref{tab.po.dolphins} and Table \ref{tab.po.Zachary}.
\begin{table}[!ht]
\caption{
the equivalence classes of the dolphins network}
\centering \(\begin{array} {|p{30pt}|p{150pt}||p{30pt}|p{150pt}|} \hline Rank&
Nodes   &Rank&
Nodes  \\
\hline 1&
2 15 37 38 41 & 2&
8 18 21 30 34 46 52 \\
\hline 3&
9 14 29 39 40 44 51 55 58 & 4&
1 16 19 24 53 60 \\
\hline 5&
7 10 22 31 35 43 48 & 6&
11 17 25 28 33 42 \\
\hline 7&
20 45 62 & 8&
3 4 6 \\
\hline 9&
26 27 56 & 10&
13 47 54 \\
\hline 11&
5 12 50 57 & 12&
36 59 \\
\hline 13&
23 32 & 14&
49 \\
\hline 15&
61 & & \\
\hline
\end{array}\)
\label{tab.po.dolphins}
\end{table}

From Table \ref{tab.po.dolphins}, we can see that the nodes no. 2, 15, 37, 38 and 41 are the most important. Here
node 37 is the most important node between two communities, node 2 belongs to the small community and the nodes no. 15, 38 and 41 belong to the big community. Here this result shows that the equivalence classes approach
can obtain representative nodes.

\begin{table}[!ht]
\caption{
The equivalence classes of the Zachary network}
\centering \(\begin{array} {|p{30pt}|p{150pt}||p{30pt}|p{150pt}|} \hline Rank&
Nodes   & Rank&
Nodes  \\
\hline 1&
1 34 & 2&
3 33 \\
\hline 3&
2 9 32 & 4&
4 14 \\
\hline 5&
6 7 20 24 28 31 & 6&
8 26 29 30 \\
\hline 7&
5 10 11 15 16 19 21 23 25 & 8&
18 22 \\
\hline 9&
13 & 10&
12 27 \\
\hline 11&
17 & & \\
\hline
\end{array}\)
\label{tab.po.Zachary}
\end{table}

From Table \ref{tab.po.Zachary}, we can see that nodes no. 1(Mr. Hi) and 34 (the chief administrator) are the
most important, because they are the kernel nodes in the separated clubs.

All the results show that the equivalence classes approach can identify the most important nodes.

\subsection{The Effectiveness of Compared Algorithms}

PageRank and HITS are two famous algorithms to sort/rank Web documents by importance. Web documents are linked by the hyperlinks and construct a directed network. Here, they are used to deal with the undirected network, the same as that the other researchers did. We calculate their effectiveness with the proposed framework to check this kind of applications. Moreover, to compare their
results to four chosen basic measure indicators, i.e., degree, betweenness, closeness, neighbors, we compute the
coverage values of these four indicators.

When computing the importance score by PageRank and HITS, we set the transition probability as 0.15 and the
iteration times as 200 for PageRank, the maximum of iteration times as 500 for HITS. The results here is obtained by the tool
jung\cite{78}.

\begin{table}[!ht]
\caption{The effectiveness of PageRank and HITS}
\centering \(\begin{array} {|p{65pt}|p{45pt}|p{50pt}|p{50pt}|p{50pt}|} \hline \multicolumn{2}{|p{70pt}|}{} &
Metabolic& Dolphins&
Zachary \\
\hline
\raisebox{-1.50ex}[0cm][0cm]{PageRank}&
Best&
0.814721&
0.868226&
0.885081 \\
\cline{2-5}
 &
Worst&
0.814721&
0.868226&
0.885081 \\
\cline{2-5}
 &
Certratio& 0.000000 & 0.000000&
0.000000 \\
\hline \raisebox{-1.50ex}[0cm][0cm]{HITS}& Best& 0.809645& 0.759840&
0.895161 \\
\cline{2-5}
 &
Worst&
0.809645&
0.759840&
0.895161 \\
\cline{2-5}
 &
Certratio& 0.000000 & 0.000000&
0.000000 \\
\hline \raisebox{-1.50ex}[0cm][0cm]{Degree}& Best& 0.984772& 0.926982&
0.991935 \\
\cline{2-5}
 &
Worst&
0.723350&
0.848260&
0.866935 \\
\cline{2-5}
 &
Certratio& 0.261422 & 0.078722&
0.125000 \\
\hline \raisebox{-1.50ex}[0cm][0cm]{Betweenness}& Best& 0.977157& 0.937250&
0.977823 \\
\cline{2-5}
 &
Worst&
0.878173&
0.918426&
0.868952 \\
\cline{2-5}
 &
Certratio& 0.098984 & 0.018824&
0.108871 \\
\hline \raisebox{-1.50ex}[0cm][0cm]{Closeness}& Best& 0.829949& 0.881346&
0.943548 \\
\cline{2-5}
 &
Worst&
0.819797&
0.869937&
0.913306 \\
\cline{2-5}
 &
Certratio& 0.010152 & 0.011409&
0.030242 \\
\hline \raisebox{-1.50ex}[0cm][0cm]{Neighbors}& Best& 0.890863& 0.879064&
0.907258 \\
\cline{2-5}
 &
Worst&
0.832487&
0.858528&
0.899194 \\
\cline{2-5}
 &
Certratio& 0.058376 & 0.020536&
0.008064 \\
\hline
\end{array}\)
\label{tab.effectiveness}
\end{table}

From Table \ref{tab.effectiveness}, we can see that all the certainty ratios of PageRank and HITS are equal to zero,
meaning that both algorithms can successfully determine the nodes with equivalent importance and do not disregard
the inequivalent nodes as equivalent.

Moreover, HITS outperforms PageRank in the Zachary network, but fails in the metabolic network and the dolphins network.

From Table \ref{tab.effectiveness}, we can see that both algorithms outperform the degree indicator. Moreover, the closeness indicator outperforms both the algorithms in all the best
coverage values and the worst coverage values. 
HITS gets a worst score in the dolphins network and PageRank gets a worst score in the metabolic network.

In general, as to PageRank and HITS, they get scores of about 76\% - 90\% for similarity ratio, and 100\% for
certainty ratio. This proves their good effectiveness, although they were not designed for the undirected networks.

The statistic results need an explanation in details. Here we choose the metabolic network as the anatomic
object. The experimental results can illustrate the preferences of both the algorithms.

Here we list the results of both two algorithm in Table \ref{tab.pagerankmetab} and
\ref{tab.hitsmetab}.

\begin{table}[!ht]
\caption{The result of PageRank for the metabolic network}
\centering \(\begin{array} {|p{24pt}|p{36pt}||p{24pt}|p{36pt}||p{24pt}|p{36pt}|} \hline Rank&Nodes& Rank&Nodes&
Rank&
Nodes \\
\hline 1& 9& 9& 2& 17&
23 \\
\hline 2& 21& 10& 3& 18&
1 \\
\hline 3& 26& 11& 27 29& 19&
28 30 \\
\hline 4& 5& 12& 13& 20&
4 \\
\hline 5& 7& 13& 12 14& 21&
6 \\
\hline 6& 19& 14& 11 15& 22&
18 \\
\hline 7& 25& 15& 16 10& 23&
8 17 \\
\hline 8& 22& 16& 24& 24&
20 \\
\hline
\end{array}\)
\label{tab.pagerankmetab}
\end{table}

From Table \ref{tab.pagerankmetab} we can see that PageRank prefers to the degree indicator and marginal
nodes. For examples, though node 7 is more important than node 5 in all aspects, PageRank sets a worst
ranking value; the relationships among the nodes no. 1, 4, 6, 8, 17 and 20 clearly demonstrate the marginal preference.  This
is the reason that PageRank does not obtain higher scores.

\begin{table}[!ht]
\caption{The result of HITS for the metabolic network}
\centering \(\begin{array} {|p{24pt}|p{36pt}||p{24pt}|p{36pt}||p{24pt}|p{36pt}|} \hline Rank&Nodes& Rank&Nodes&
Rank&
Nodes \\
\hline 1& 9& 9& 23& 17& 15 11 \\
\hline 2& 21& 10& 8 17& 18&
3 \\
\hline 3& 22& 11& 20& 19&
4 \\
\hline 4& 19& 12& 24& 20&
12 14 \\
\hline 5& 25& 13& 18& 21&
28 30 \\
\hline 6& 26& 14& 5& 22&
13 \\
\hline 7& 7& 15& 27 29& 23&
2 \\
\hline 8& 16 10& 16& 6& 24&
1 \\
\hline
\end{array}\)
\label{tab.hitsmetab}
\end{table}

From Table \ref{tab.hitsmetab}, we can see that HITS prefers to the neighbors and closeness. For examples, in
the results of HITS, node 8 and node 17 are more important than node 5, node 25 is more
important than node 26. These preferences would lead to a result that important bridge and small
community may be somehow ignored. The bad score of HITS on the dolphins network also can be explained on these
preferences.

The bias of HITS can be explained by itself. HITS determines two values for a node: the authority and the hub value, which are mutually recursively defined. The authority value of a node is the sum of the scaled hub values that point to this node. The hub value of a node is the sum of the scaled authority values of the nodes that this node points to. When all the links are two-way, because of the iterations, the central nodes are emphasized and certainly, the nodes connecting to the central nodes are also emphasized.

Generally speaking, PageRank and HITS are both good to identify the node importance, but we must be aware of their bias.

\section{The Internet Structure}

Based on BGP tables posted at archive.routeviews.org, Mark Newman reconstructed a network representing the structure of the Internet\footnote{The file name is as-22july06.zip, which can be found in the website of Newman.}. In this network, each node represents a domain and each edge represents an inter-domain interconnection. This network has 22963 nodes and 48436 edges.

This paper uses the equivalence classes approach to analyze the Internet, and lists the top 10 equivalence classes
 as Table \ref{tab.po.internet}.

\begin{table}[!ht]
\caption{The top 10 equivalence classes of the Internet}
\centering \(\begin{array} {|p{30pt}|p{300pt}|} \hline Rank & Nodes  \\
\hline 1 & 4 15 23 27\\
\hline 2 & 3 11 39 40 59\\
\hline 3 & 7 12 36 51 55\\
\hline 4 & 16 20 25 43 56 64 128\\
\hline 5 & 14 38 42 129 158\\
\hline 6 & 21 46 53 58 1282 1752\\
\hline 7 & 1 13 18 24 35 99 296 1271 1868\\
\hline 8 & 19 26 32 37 45 61 1761 2363 2910\\
\hline 9 & 5 28 29 63 157 189 219 1272 1279 1826 1895 2493\\
\hline 10 & 22 69 161 180 333 1454 1497 1833 1875\\
\hline
\end{array}\)
\label{tab.po.internet}
\end{table}

For Table \ref{tab.po.internet} we can see that the most important nodes are no. 4, 15, 23, 27.

\begin{figure}[htbp]
  \subfigure[the top layers]{
    \label{fig:subfig:a} 
    \includegraphics[width=7.2cm,height=5.76cm]{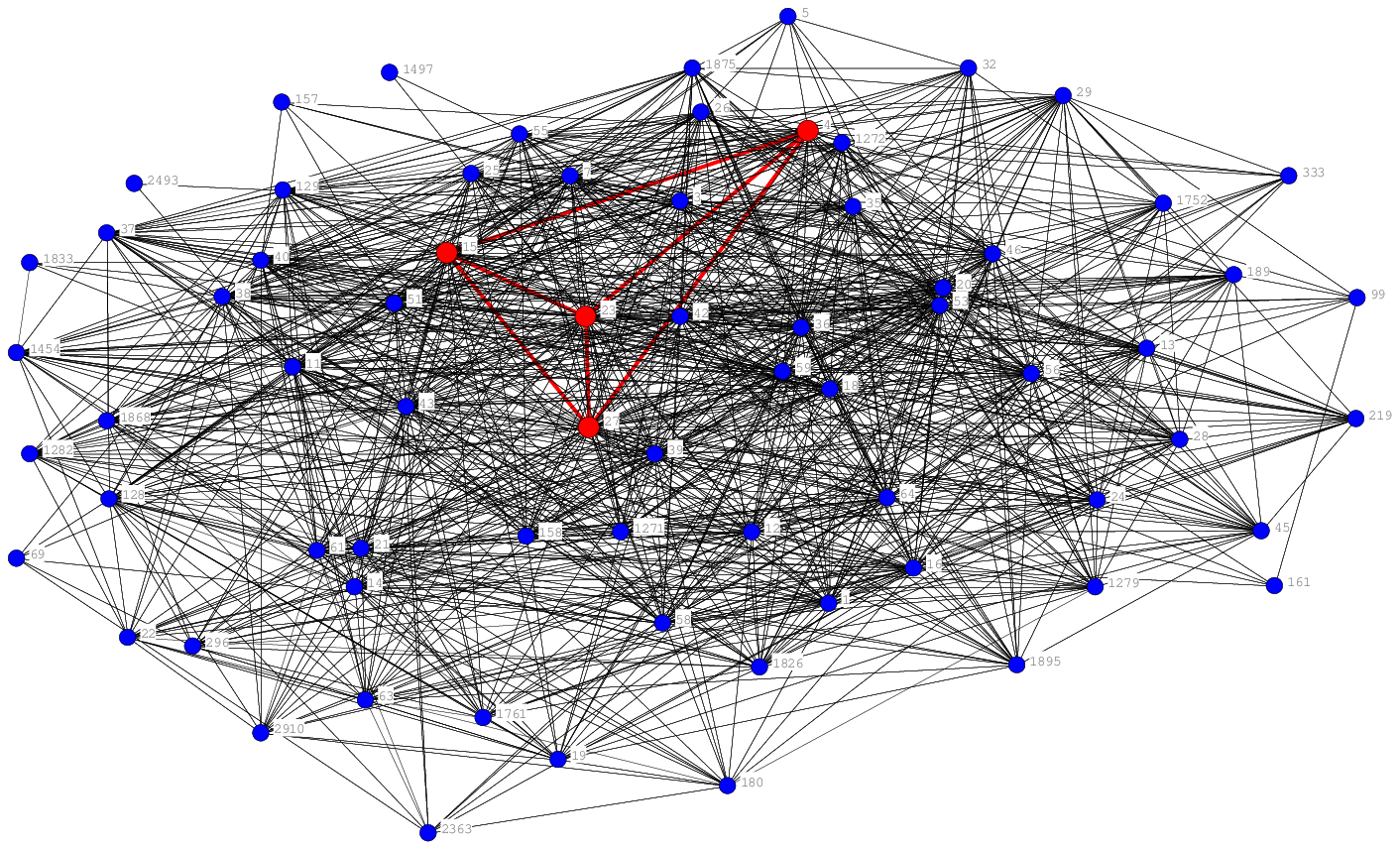}}
  \hspace{-0.5cm}
  \subfigure[the top two layers]{
    \label{fig:subfig:b} 
    \includegraphics[width=7.2cm,height=5.76cm]{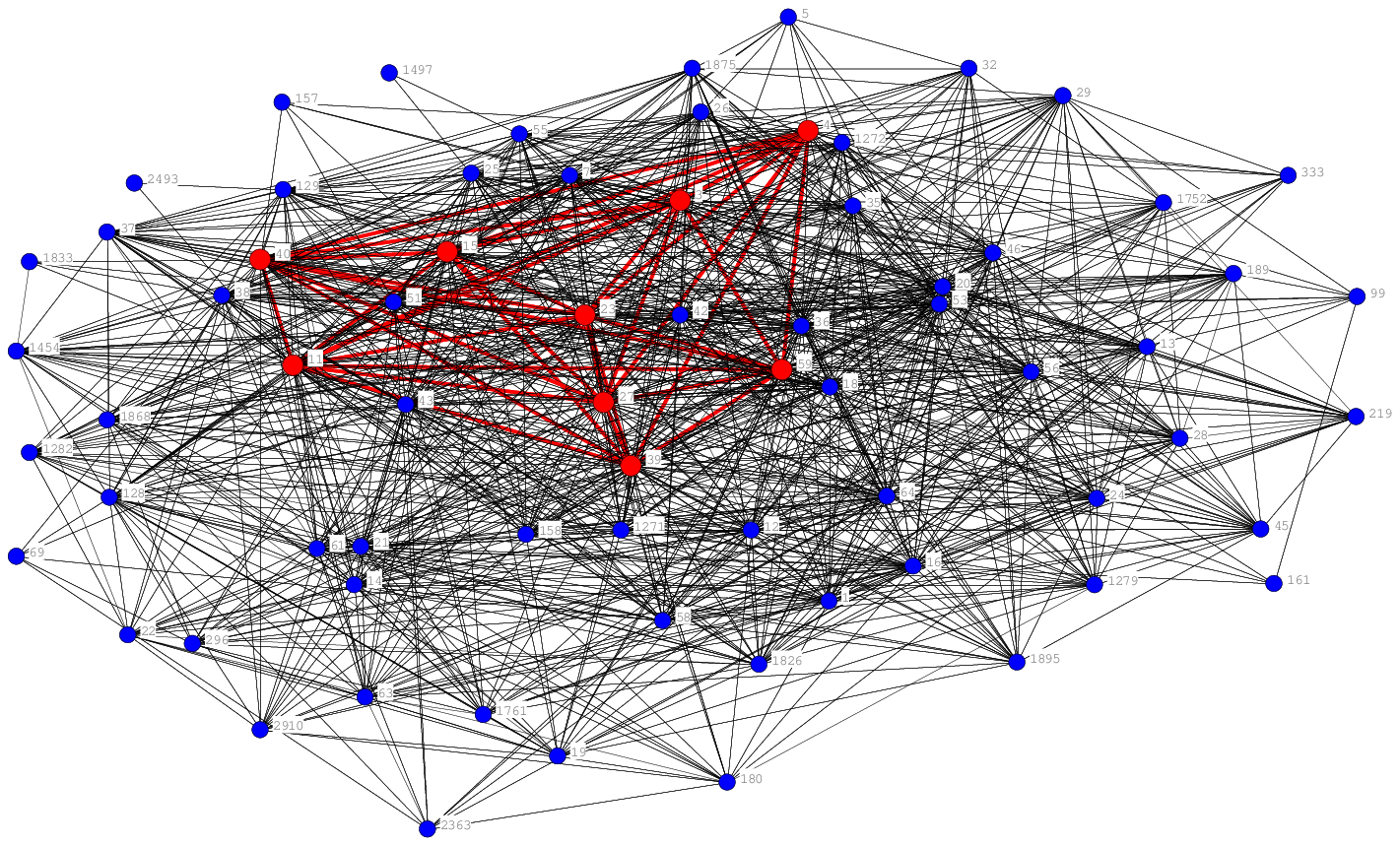}}
  \caption{The Internet kernel with 71 hub nodes and 1102 edges.}
  \label{fig.internet} 
\end{figure}

As we have known, the Internet is a scale-free network\cite{221}. A few hub nodes have a great amount of links, and most nodes have a few links. The Internet also have the rich-club phenomenon\cite{222}, that is, the hub nodes tend to link to the others. This paper retracts a sub-network which only contain the nodes in Table \ref{tab.po.internet} with the proposed approach, and the result demonstrates that the Internet has the rich-club phenomenon with the visualization technology as Fig. \ref{fig.internet}.

From Fig. \ref{fig.internet}, the average degree of this subnetwork is 31.0423, contrarily, the whole network is 2.1093, meaning that the hub nodes forms a dense kernel. From Fig. \ref{fig:subfig:a}, the most important 4 nodes are red, and they form a complete graph.
Moreover, the proposed approach can demonstrate the hierarchy of the Internet. As shown in Fig. \ref{fig:subfig:b}, the red nodes form a very dense graph, however, the graph lacks some links to become a complete graph, that is, if the node is more important, it is more likely linked by the kernel.
%

\section{Conclusions}
This paper proposes a framework to investigate the node importance. The proposed framework suggests four
rules to characterize the node importance and suggests the use of the equivalence classes to describe the
relationship among nodes because of the desired features of the equivalence classes approach. This paper also suggests that the equivalence classes
approach can be used as a benchmark to measure the effectiveness of the other ranking/sorting algorithms, and
proposes three sub-indicators based on the coverage indicator.

This paper demonstrates how to apply this framework. Three real-world networks are used to calculate the
effectiveness of PageRank and HITS, and the experimental results show that both algorithms perform well. Moreover, the
analysis on the metabolic network showed that: PageRank would wrongly order some nodes because of the bias on
degree and marginal nodes; HITS has the biases on
neighbors and closeness. These results imply that we must be aware of the bias of PageRank and HITS when using them as a benchmark in such a field. On the other hand, considering the scores of both the algorithms,
they may be capable of being used as rapid algorithms on this issue. From the experimental
results we can see that this framework is feasible.

Moreover, this paper applies the proposed approach to analyze the Internet. The experimental results show that the Internet has a kernel with dense links and have the rich-club phenomenon with the computer visualization technologies.

In the future, we will extend it to more complicated cases, for example, the directed and weighted networks. The proposed method is potential to be applied into the analysis of critical proteins and find the drug targets and other fields.

\section{Acknowledgments}
The authors thank for UCINET and jung for the convenience. The authors gratefully thank the editors and
the anonymous reviewers for the improvement of the quality of this paper.


\end{document}